# Strong enhancement of magnetic ordering temperature and structural/valence transitions in EuPd$_3$S$_4$ under high pressure

Shuyuan Huyan[1,2], Dominic H. Ryan[3], Tyler J. Slade[1,2], Barbara Lavina[4,5], Greeshma C. Jose[6], Haozhe Wang[7], John M. Wilde[1,2], Raquel A. Ribeiro[1,2], Jiyong Zhao[5], Weiwei Xie[7], Wenli Bi[6], Esen E. Alp[5], Sergey L. Bud'ko[1,2,*], Paul C. Canfield[1,2,*]

[1] *Ames National Laboratory, US DOE, Iowa State University, Ames, Iowa 50011, USA*

[2] *Department of Physics and Astronomy, Iowa State University, Ames, Iowa 50011, USA*

[3] *Physics Department and Centre for the Physics of Materials, McGill University, 3600 University Street, Montreal, Quebec, H3A 2T8, Canada*

[4] *Center for Advanced Radiation Sources, The University of Chicago, Chicago, IL 60637, USA*

[5] *Advanced Photon Source, Argonne National Laboratory, Argonne, IL 60439, USA*

[6] *Department of Physics, University of Alabama at Birmingham, Birmingham, AL 35294, USA*

[7] *Department of Chemistry, Michigan State University, 578 S Shaw Lane, East Lansing, MI 48824, USA*

\* Sergey L. Bud'ko

Email:  budko@ameslab.gov

\* Paul C. Canfield

Email:  canfield@ameslab.gov

## Abstract

We present a comprehensive study of the mixed valent compound, EuPd$_3$S$_4$, by electrical transport, X-ray diffraction, time-domain $^{151}$Eu synchrotron Mössbauer spectroscopy, and X-ray absorption spectroscopy measurements under high pressure. The electrical transport measurements show that the antiferromagnetic ordering temperature, $T_N$, increases rapidly from 2.8 K at ambient pressure to 23.5 K at ~19 GPa and plateaus between ~19 and ~29 GPa after which no anomaly associated with $T_N$ is detected. A pressure-induced first order structural transition from cubic to tetragonal is observed, with a rather broad coexistence region (~20 GPa to ~32 GPa) that corresponds to the $T_N$ plateau. Mössbauer spectroscopy measurements show a clear valence transition from approximately 50:50 $Eu^{2+}$:$Eu^{3+}$ to fully $Eu^{3+}$ at ~28 GPa, consistent with the vanishing of the magnetic order at the same pressure. X-ray absorption data show a transition to a fully trivalent state at a similar pressure. Our results show that pressure first greatly enhances $T_N$, most likely via enhanced hybridization between the Eu 4f states and the conduction band, and then, second, causes a structural phase transition that coincides with the conversion of the europium to a fully trivalent state.

**Main Text**

**Introduction**

In general, mixed valency in f-electron systems occurs when the more-localized f-orbitals of rare-earth elements hybridize with s, p, and/or d electrons. A quantum superposition of different f-orbital valences emerges when the energy difference between competing single-valent states is smaller than the f-electron bandwidth, namely fluctuating valence state [1]. The onset of mixed valence under external pressure, chemical substitutions, or thermal contraction has dramatic consequences on the macroscopic properties of f-electron systems including lattice collapse [2], quenched magnetism [3], superconductivity [4,5], Kondo behavior [6], and quantum criticality [4,5]. Despite being central to f-electron physics, the underlying mechanism of mixed-valence phenomena is still not fully understood due to the experimental limitations and the lack of ideal materials.

That said, the Eu intermediate valence behavior in $EuPd_3S_4$ is rather special. The rare-earth palladium bronzes, $RPd_3S_4$, were reported to crystallize in the cubic $NaPt_3O_4$ structure ($Pm\bar{3}n$, SG #223) with the rare earth occupying the 2a site and forming a bcc sublattice [7]. Whereas the $RPd_3S_4$ compounds exist for all the trivalent rare earths, they do not appear to form with the divalent alkaline earths (Ca, Sr,…). Interestingly, when prepared with europium [7] or ytterbium [8] a roughly 50:50 mix of divalent and trivalent rare earth was found, which makes the intermediate valence in these two compounds distinct from the more general fluctuating valence phenomenon case. In $EuPd_3S_4$, 50 % $Eu^{2+}$ makes the material magnetic as observed in Mössbauer spectroscopy and thermodynamic measurements [7,9]. By chemical substitution, smaller $Y^{3+}$ ions tend to replace $Eu^{3+}$ sites first and make the valence of Eu more divalent; however, substitution with bigger $La^{3+}$ ions on the other hand, promotes the electron hopping between $Eu^{2+}$ and $Eu^{3+}$, leading to an intermediate Eu valence until 60% of Eu have been replaced, above which, Eu becomes more trivalent. The chemical substitution work suggests that preservation of the unit cell volume size is a dominant factor controlling the $Eu^{2+}$: $Eu^{3+}$ valence ratio [9]. Very recently, it was found that rather than adopting the $Pm\bar{3}n$ structure with a single, crystallographic Eu site (2a) hosting the two Eu valences, $EuPd_3S_4$ was found to take the more reasonable cubic $Pm\bar{3}$ (SG #200) structure with two distinct Eu sites (1a and 1b) below 340 K at ambient pressure [10].

Considering that chemical substitution and the associated changes in unit cell were shown to profoundly impact the mixed valence state in $EuPd_3S_4$, it is worthwhile to appreciate that the effect of external pressure has proven to be an effective way to tune the atomic distances and thus may be used to increase the degree of orbital hybridization, and is considered to be a cleaner and more powerful tool for modifying the valence state than chemical doping. Here, we report the systematic high-pressure investigation of $EuPd_3S_4$ using a combined experimental approach including electrical transport measurements, X-ray diffraction (XRD), time-domain synchrotron Mössbauer spectroscopy (SMS), and partial fluorescence-yield X-ray absorption spectroscopy (PFY-XAS). A significant, about 8-fold, enhancement in antiferromagnetic transition temperature, ($T_N$) from ~2.8 K at ambient pressure [7] to about 23.5 K at ~19 GPa is observed. Subsequently, in a large pressure window from ~19 GPa to ~29 GPa $T_N$ is almost pressure independent, whereas the size of resistance drop below $T_N$ is gradually suppressed with the pressure. The sample in this pressure window is found to be in a mixed phase of cubic and tetragonal structures with the ratio of cubic/tetragonal decreasing with pressure. Above ~29 GPa, an abrupt disappearance of antiferromagnetic (AFM) order is observed. In the same pressure range, both SMS and XAS measurements show a clear change from 50:50 $Eu^{2+}$:$Eu^{3+}$ to fully $Eu^{3+}$, and is coincident with the complete disappearance of the lower pressure, cubic phase, resulting in a single phase of the tetragonal strcuture. This comprehensive study beautifully illustrates the interplay of crystal structure, magnetic ground state, and the associated valence state tuned by high pressure, and suggests that $EuPd_3S_4$ provides a clean and simple system for the detailed study of how magnetism, valence and structure are intertwined and should serve as a benchmark for theoretical efforts to model and understand mixed-valence behavior.

**Results**

Fig. 1(a) presents the temperature dependence of the resistance of single crystalline EuPd$_3$S$_4$ under pressure up to ~32 GPa. The overall R-T behavior is metallic over the whole temperature and pressure ranges covered here. A clear resistance drop is seen below 2.8 K at 0.4 GPa, reflecting the loss of spin disorder scattering as AFM ordering occurs. We define $T_N$ as the mid-point of the step in the temperature derivative of resistance (dR/dT), with the uncertainty taken as half of the step width, as shown in Fig. 1(b). $T_N$ at 0.4 GPa is consistent with the ambient pressure heat capacity and magnetization results reported previously [7,9]. Increasing the pressure from 0.4 GPa to ~19 GPa leads to a marked rise in $T_N$ from 2.8 K to 23.5 K, Fig. 1(d). The pressure dependence of the resistance for pressure up to ~19 GPa is modest and without any clear features, as shown in Fig. 1(c). Above 20 GPa, there is a sharp increase in resistance; the position of the $T_N$ step remains essentially unchanged, but the size of the step-like feature in dR/dT becomes progressively smaller and is lost by ~29 GPa (Fig. 1(b)). As will be seen below, when we examine the powder x-ray data, these changes are consistent with the observation that for ~19-29 GPa, the system passes through a crystallographic two-phase region as the increasing pressure progressively converts the AFM-ordering Eu$^{2+}$-containing phase into a fully trivalent non-magnetic phase. Above ~29 GPa, the features marking loss of spin-disorder scattering in the resistance, and the step like feature in dR/dT are gone, and no further evidence for magnetic ordering (at least above 1.8K, our lowest measurement temperature) is observed, suggesting that the conversion is now complete. Of note is that the shape of R-T curve clearly changes from concave-like at low pressures to convex-like at high pressures, indicating possible change of band structure and/or paramagnetic and phonon scattering.

In addition to the step like feature that is related to $T_N$, a slope change in dR/dT at lower temperature (marked as T*, in Fig. 1(b)) is also observable at and above 12 GPa and becomes clearer as pressure increases to ~19 GPa (Fig. 1(b)) Similar features in dR/dT have been reported previously in GdB$_4$ [27] and SmB$_4$ [28] and were proposed to reflect an intrinsic bulk property based on heat capacity results [27]. On the other hand, extra terms that involve electron-magnon scattering above a certain temperature could also cause such a dR/dT anomaly [29]. Although the origin of this feature is not fully understood and still needs more study, it appears to be secondary to the clear feature associated with $T_N$.

The PM-AFM transition temperature, $T_N$, as a function of the pressure for EuPd$_3$S$_4$ is presented in Fig. 1(d). We can see that with increasing pressure, $T_N$ has an approximately eight-fold enhancement from 2.8 K at 0.4 GPa to around 23.5 K at ~19 GPa with an average increasing rate of ~1.1 K/GPa, then $T_N$ remains essentially constant from ~19 GPa to ~29 GPa, above which signatures of antiferromagnetic ordering are not observed. Meanwhile, resistance versus pressure (Fig. 1(c)) at all temperatures shows an abrupt increase starting from ~22 GPa. This is most likely due to the onset of a crystallographic two-phase region (as will be discussed below) but may also be associated with the commensurate change in valence and therefore band structure. As will be discussed further below, the flat $T_N$ versus P region between ~19 GPa and ~29 GPa is most likely the result of entering a region of mixed crystallographic phases with increasing pressure leading to a larger fraction of the smaller volume, purely Eu$^{3+}$ phase. As a result of this two-phase nature, we speculate that the strain on the remaining Eu$^{2+}$ phase remains effectively constant over this pressure range, leading to an effectively constant $T_N$ value.

It is notable that the residual resistance ratio (RRR = R(300 K)/R(1.8 K)) as a function of the pressure shows a different behavior compared with the resistance as a function of the pressure, as shown in Fig. 1(c). The RRR first increases from ~40 at 0.4 GPa to ~100 at ~10 GPa and then rapidly decreases to less than 2 at ~22 GPa, above which the resistance at all temperatures shows a sharp increase with the pressure, which might indicate a structural transition.

The synchronous change of resistance at all temperatures from 1.8 K to 300 K, as shown in Fig. 1(c) suggests that if there is any structural/valence transition under pressure, the transition temperature should be above 300 K and its phase line is (almost) vertical in temperature-pressure (T-P) plot. Therefore, it will be sufficient to conduct measurements at room temperature to address the possible structural/valence transition.

To explicitly explore the structural stability of EuPd$_3$S$_4$ under pressure, synchrotron X-ray diffraction experiments were performed on a sintered powder (polycrystalline) sample up to ~57 GPa. Figs. 2(a-c) demonstrates the outcomes. Peaks from copper (calibrant) and rhenium (gasket) are indexed in Figs. 2(b) and 2(c) and are marked with orange stars and purple crosses, respectively, in Fig. S1 to illustrate their systematic shift with pressure. Three unique pressure regions were detected, as shown in Fig. 2(a): zone I (ambient to ~17 GPa), zone II (~20 GPa to ~32 GPa), and zone III (~34 GPa to ~57 GPa), respectively. The diffraction patterns in zone I can be indexed to the cubic space group $Pm\bar{3}$ (#200), as shown in Fig. 2(b), which is consistent with the structure of the specimen at ambient pressure [10]. In zone III, the patterns are resolvably different and could be assigned to a tetragonal space group, P4/mmm (#123), as depicted in Fig. 2(c). The pressure dependences of the lattice parameters and unit cell volume $EuPd_3S_4$ are shown in Figs. 2(d) and 2(e). Clear discontinuous changes in both lattice parameters (a and c) and the unit cell volume (V) are observed. Considering that P4/mmm (#123) is not a translationengleiche subgroup of $Pm\bar{3}$(200), which is actually the translationengleiche subgroup of $Pm\bar{3}n$(223), we expect that the structural transition under pressure to be first order according to Hermann's theory [30,31]. The pressure-volume (P-V) curves were analyzed using a second-order Birch-Murnaghan equation of state (BM EOS) over the two separate pressure ranges shown in Fig. 2(e). At pressures up to ~18 GPa (zone I), the elastic parameters yield a zero-pressure bulk modulus $B_0$ of 103.5(20) GPa and a zero-pressure unit-cell volume $V_0$ of 297.89 Å$^3$ at ambient conditions which is consistent with the reported value [9]. The unit cell volume in zone III shows a clear deviation from the fitted curve based on the data points in zone I with the unit cell volume at ~34 GPa decreasing by ~ 4.8%, which is intricately linked to the structural transition and the substantial change in the valence of Eu ions. It is worth mentioning that the fitted parameters in zone I were also obtained from a larger volume of data taken from multiple experimental runs, as shown in Fig. S5.

As diffraction peaks from both the cubic and tetragonal phases are observed in zone II, a two-phase scenario is a reasonable interpretation. The gradual disappearance of the cubic phase and the appearance and growth of the tetragonal phase can be seen clearly despite the increasing difficulty of indexing the diffraction patterns within zone II. Fig. 3 shows the changes of the specific diffraction peaks in detail, highlighting the disappearance of the cubic (210) and (211) Bragg peaks and the sudden emergence of the tetragonal (113) peak at ~20 GPa. Figs. S2-S4 show more detail.

A first-order structural phase transition typically involves a sudden change in lattice parameters and unit cell volume, and the presence of a two-phase region in the intermediate pressure zone. Fig. 2(d) and 2(e) illustrate all of these features clearly. However, the large two-phase window, more than 10 GPa, cannot simply be attributed to pressure gradients within the pressure transmitting medium since the pressure gradient in neon at 20 GPa is only 0.15 GPa [17], which is almost two orders of magnitude smaller than the observed two-phase pressure range. The broad two-phase region must therefore be considered to be an intrinsic feature associated with the valence change in $EuPd_3S_4$ under these conditions. An analogous large two-phase pressure range has also been reported in SmSe, SmTe [32], and $CaFe_2As_2$ [33,34], possibly as a result of anharmonic strain effects caused by the significant volume mismatch between the phases [35]. Noteworthy, the Néel temperature plateau in Fig. 1 (d), where $T_N$ is almost independent of pressure, coincides well with the two-phase pressure region that has a mixed phase with shifting phase ratio. In other words, the behavior of the $T_N$ plateau and the gradual loss of magnetic moment in the $T_N$ plateau region are likewise intrinsic and may be closely related to the gradual increase in the relative amount of the tetragonal phase, and, as will be shown below, non-magnetic-$Eu^{3+}$.

The conventional, energy-domain, $^{151}$Eu Mössbauer spectrum of $EuPd_3S_4$ shows two, roughly equal-area, well-separated lines from the $Eu^{2+}$ and $Eu^{3+}$ present [7,9]. These two lines reflect the two different transition energies (i.e., isomer shifts), and hence photon frequencies, associated with the two europium valence states. In the time-domain SMS spectra we observe a simple beat pattern between these two photon frequencies (see Fig. 4(a) at 0.7 GPa) with the beat frequency set by the fractional difference in photon energies (or frequencies), and the amplitude set largely by the relative populations of the two valence states. A visual inspection of Fig. 4(a) reveals that both change with increasing pressure: the beat frequency decreases, and the amplitude modulation generally becomes less distinct. The evolution in beat frequency indicates that the isomer shift difference between the two components is becoming smaller. Finally, there is an abrupt change in the curves above 27 GPa to a relatively very long beat period (rather than a simple exponential decay that we would associate with a single trivalent line) suggesting that the system is not just the cubic crystallographic phase with pure $Eu^{3+}$.

More intuitive energy-domain spectra generated by CONUSS [22] by fitting the time-domain data are shown in Fig. 4(b). Within the pressure range of 0.7 GPa to 27 GPa, two distinct peaks are observed, representing the isomer shifts of $Eu^{2+}$ and $Eu^{3+}$ at low pressures. For clarity, these two peaks are referred to as the $Eu^{2+}$ and $Eu^{3+}$ peaks, although it should be noted that the oxidation state of the $Eu^{2+}$ peak tends to approach that of $Eu^{3+}$ under pressure. The intensity of the $Eu^{2+}$ peak weakens around 27 GPa and completely diminishes at 29.7 GPa, resulting in the sole presence of the $Eu^{3+}$ peak, thereby indicating a complete transition from an initial nearly 50:50 $Eu^{2+}$:$Eu^{3+}$ state to a fully $Eu^{3+}$ state. Concurrently, the linewidth of $Eu^{3+}$ peak exhibits a clear broadening at 29.7 GPa compared to lower pressures. This can be attributed to the presence of a quadrupole interaction, $\Delta E(Q)$, of 2.16(5) mm/s, resulting from the emission of a gamma ray from $^{151}$Eu during the transition between the excited state (spin 7/2) and the ground state (spin 5/2). This observation is consistent with the structural transition from a cubic to a tetragonal phase, where an electrical field gradient tensor arises. However, due to the crowded arrangement of multiple components split by the quadrupole interaction within the line width of ~2.3 mm/s, the absorption spectrum is not resolvable. Besides, it is noteworthy that the quadrupole interaction, ΔE(Q), roughly increases within the pressure range of approximately 28 to 30 GPa, as shown in Fig. S7. This increase can be attributed to the increase of structural anisotropy. The observed enhancement in the quadrupole interaction may correspond to the transition from a two-phase region, characterized by the presence of strain between the two phases, to a fully tetragonal phase. Upon decompression, the absorption peak corresponding to $Eu^{2+}$ recovers at ~25 GPa, indicating the reversibility of the pressure.

The isomer shift exhibited the same evolution with pressure in a second experimental run on a sintered powder sample (SMS data are shown in Fig. S6) as it did for a single crystal sample. Absolute calibration of the isomer shifts of the two components was achieved by measuring SMS spectra at several pressures after introducing a standard (EuS) with a known isomer shift (-11.496 mm/s) in the beam after the pressure cell (detailed explanation of calibration of absolute values of isomer shifts is shown in Fig. S8, S9). These corrected measurements allowed us to show that most of the isomer shift change is due to the $Eu^{2+}$ line moving to more positive values (Fig. 4(c)) with increasing pressure.

Variations were observed in the area fraction of the $Eu^{2+}$ peak within the two-phase region (approximately 21 to 27 GPa, as shown in Fig. 4(d)). These differences may stem from slight variations between the single crystal and powder samples or the effects of structural instability within the two-phase region, where both cubic and tetragonal phases coexist.

The pressure dependence of the mean Eu valence was estimated by assuming that the area of each component in the SMS spectrum was proportional to the number of ions associated with the corresponding valence state. The detailed calculation is shown in Fig. S10, and the result is summarized in Fig. 6(d), shown below.

PFY-XAS measurements were performed to provide independent confirmation of the valence change. As shown in Fig. 5(a), the absorption peak corresponding to $Eu^{2+}$ gradually weakens as the pressure increases to 14 GPa, then becomes much weaker at ~22 GPa, and remains as a small bump for higher pressures, clearly demonstrating a transition from 50:50 $Eu^{2+}:Eu^{3+}$ to essentially $Eu^{3+}$. The valence transition appears at a somewhat lower pressure in the PFY-XAS measurements than in the SMS results. Such a shift may be associated with non-hydrostaticity due to the lack of a PTM in the PFY-XAS measurement. The persistence of a tiny $Eu^{2+}$ peak at higher pressures may be attributed to pressure inhomogeneity. Furthermore, a small, but discernible shift towards slightly lower energy of the absorption peak corresponding to $Eu^{3+}$ is noted at and above ~22 GPa. The possible reasons for this tiny shift in the absorption peak position, aside from experimental factors such as energy resolution and XAS instrument calibration, could be a change in the crystal field, which splits the 4f electronic state into different energy levels. This shift could be closely related to the structural transition that begins to occur between ~20 GPa and ~32 GPa. To fully comprehend the mechanism behind the absorption peak shift, additional research, such as DFT calculation based on the more accurate structure, is required.

Modeling the PFY-XAS data using a series of Lorentzian and arctangent functions for each absorption peak, as seen in Fig. 5(b), yields the mean valence of the Eu ion. The average valence is estimated using the following formula:

$$v = \frac{2 \times A_{Eu^2} + 3 \times A_{Eu^3}}{A_{Eu^2} + A_{Eu^3}} \qquad (1)$$

where $A_{Eu^2}$ and $A_{Eu^3}$ are the areas of absorption peaks for $Eu^{2+}$ and $Eu^{3+}$, respectively. The estimated mean Eu valence as a function of pressure as measured by SMS and XAS are depicted in Fig 6(d) and (Fig 5(c) & Fig 6(e)), respectively. Both measurements demonstrate a clear valence change from $Eu^{2+}$ to $Eu^{3+}$. The shift is abrupt and occurs at higher pressure (> ~27 GPa) in the SMS measurements with He as the PTM. Due to the low data density and the fact that the high-pressure XAS measurement is conducted without PTM, it is impossible to determine the width of the valence transition; however, it is clear that the mean valence is very close to 3 at and above 24 GPa.

**Discussion**

The pressure dependence of the AFM transition temperature $T_N$, resistance at 1.8 K, the volume of the unit cell, and the mean valences estimated from SMS and XAS data is presented in Fig. 6. Two vertical lines at ~ 20 and 32 GPa mark the lower and upper boundaries of the two-phase region and are shown crossing all the data sets. Although the use of different PTMs and manometers for different high-pressure measurements made at different temperatures may cause some inconsistency in the critical pressures and smearing of the first-order transition, a clear picture still can be obtained: There are three distinct pressure regions associated with the low-pressure cubic phase, an intermediate pressure two-phase region where both cubic and tetragonal phases are present, and finally the high-pressure tetragonal phase (see Fig. 6(c)). Two pressures (~20 GPa and ~32 GPa) separate the changes we see in the T-P phase diagram shown in Fig. 6(a) as well as the R(1.8 K) data shown in Fig. 6(b). Fig. 6(a) shows that $T_N$ rises rapidly in the low-pressure region, is essentially constant in the intermediate pressure region, and is not detectable (due to the loss of moment bearing $Eu^{2+}$) in the high-pressure region. Fig. 6(b) shows that the 1.8 K resistance of the low-pressure phase is relatively low and essentially unchanging in the low pressure phase. In the intermediate pressure region, R(1.8 K) rises substantially in a near linear manner before appearing to saturate above 30 GPa. The unit cell volumes for the low-pressure and high-pressure phases are shown in Fig. 6(c). In the two-phase region, there is an increasing amount of tetragonal phase as pressure increases. The estimated mean valence derived from SMS results is shown in Fig. 6(d). The mean valence is nearly constant up to ~20 GPa, increases in a non-monotonic manner, depending upon the run and sample, over the next 10 GPa and is fully trivalent at and above 30 GPa. The mean valence derived from XAS data in Fig. 6(e) is fully consistent with the SMS data but suffers from two problems. First, the XAS data is the only data set collected without pressure medium, and as such represents very different pressure conditions; second, the XAS data set it too sparse to accurately detect an onset to valence change. All of this said, the XAS data set is qualitatively similar to the SMS data, showing a change from roughly 50:50 $Eu^{2+}$:$Eu^{3+}$ to essentially $Eu^{3+}$ by ~ 25 GPa. The fact that this pressure is lower than any seen in the other measurements is most likely associated with the less hydrostatic conditions. Taken all together, then, our diverse data sets are consistent with $EuPd_3S_4$ being in the low-pressure cubic phase up to ~ 20 GPa. In this pressure range there is a large increase in $T_N$ that we associated with an increasing coupling (or hybridization) between the Eu 4f shell and the conduction electrons. Whereas such large $dT_N/dP$ values are rather uncommon for rare earth compounds, similarly large values have been found for some Eu-based intermetallics, such as $EnIn_2As_2$ [36], $EuSn_2As_2$ [37], $EuSn_2P_2$. [38], and $EuMnBi_2$ [39].

Ultimately, when the coupling (or hybridization) between the conduction electrons and the Eu-4f-shell becomes sufficiently strong, the $EuPd_3S_4$ sample starts transforming into the high pressure, $Eu^{3+}$-tetragonal phase and enters into the two-phase region of the T-P phase diagram. As more and more of the sample transforms into the smaller volume, tetragonal phase, the remaining cubic phase $EuPd_3S_4$ remains at a more or less constant strain and constant $T_N$. Finally, near 32 GPa the conversion is complete and the sample becomes fully tetragonal and fully non-magnetic $Eu^{3+}$. The T-P phase diagram shown in Fig. 6(a) bears some resemblance to proposed "global phase diagrams" [40,41] for Eu-based compounds, but also has some key differences. At a gross level of detail, the increase of $T_N$ with pressure followed by a much higher slope (vertical in the case of $EuPd_3S_4$) transition line to the non-magnetic, trivalent state is qualitatively similar to what is proposed by earlier "global" phase diagrams. That said, there are key differences, primarily associated with the fact that we find a first-order, structural phase transition from a low-pressure cubic phase to a high-pressure tetragonal phase. This gives rise to the well-defined two-phase region that we find between ~20 and ~32 GPa and, perhaps more importantly, makes the existence of a critical end point to the structural phase transition line impossible. As such, then, $EuPd_3S_4$ presents a unique example of pressure induced, valence collapse in an Eu-based compound.

## Materials and Methods
### *Crystal growth*
High quality single crystals of $EuPd_3S_4$ were grown by a two-step solution growth method by adding Eu to a Pd-S melt [11]. First, a nominal composition of $Eu_5Pd_{58}S_{37}$ was loaded into a fritted alumina crucible set [12,13] and sealed in a fused silica tube. The tube was heated to 1150°C, held for 8 hours, and cooled over 36 hours to 1050°C, after which the liquid was decanted. The tube was opened, all solidified sulfides and oxides were discarded, and the captured decanted liquid reused in a new crucible set. The second crucible set was again sealed and then warmed to 1075°C. After holding for 8 hours, the furnace was slowly cooled over 150 hours to 900°C, at which point the remaining solution was decanted. After cooling to room temperature, the crucibles were opened to reveal large, mirror-faceted crystals [11].

The polycrystal samples (used for synchrotron X-ray diffraction, Mössbauer spectroscopy, and X-ray absorption spectroscopy measurements) were prepared by a direct solid-state reaction from a stoichiometric mixture of EuS (99.9%), Pd (99.95%) and S (99.95%) powders [9]. The mixture was pressed into a pellet, loaded in an alumina crucible, and sealed in a fused silica tube with a partial pressure of helium gas. The tube was heated to 650 °C over 3 hours, held for an hour and then heated to 900 °C over 3 hours, and held for 90 hours before furnace cooling. The resulting, sintered pellets were checked for phase purity by powder x-ray diffraction and then, as needed, ground, repressed into a pellet and heated again to 900 °C for further reaction [9].

## High pressure measurements

### Electrical transport

Linear four-terminal electrical resistivity measurements were performed in a Diamond Anvil cell (DAC) (Bjscistar, [14]), with 500 μm culet-size standard cut-type Ia diamonds. $EuPd_3S_4$ single crystals were cleaved into 20 μm thick flakes and cut and polished into 100 μm x 40 μm plates. A single plate was loaded together with a tiny ruby sphere (< 10 μm in diameter) into an apertured stainless-steel gasket covered by cubic-BN. Platinum foil was used to create electrodes to connect to the sample. Nujol mineral oil was used as pressure transmitting medium (PTM), since: 1) this fluid medium can maintain a quasi-hydrostatic pressure environment with a small pressure gradient below its liquid/glass transition [15-17]; 2) the use of a fluid medium helps to minimize direct contact between the sample and diamond culet which could lead to an uniaxial pressure component. Pressure was determined by ruby fluorescence [18] at room temperature. Low temperature resistance measurements down to 1.8 K were conducted in the Quantum Design Physical Property Measurement System (PPMS).

### X-ray diffraction (XRD)

Two room-temperature high-pressure powder XRD (PXRD) runs were carried out at the GSECARS 13-BM-D and XSD 3-ID-B Beamlines of the Advanced Photon Source (APS), at Argonne National Laboratory (ANL). X-rays with a wavelength of 0.29521 Å and 0.4833 Å were focused to a 15 μm (vertical) × 15 μm (horizontal) spot size at both beamlines. The sintered powder sample and ruby spheres (run1,2)/ Cu powder (for manometry, in run3; PXRD data shown in main text are all from run3), were loaded into a wide opening SSDAC-70 DAC, diamond anvils with 300 μm and 500 μm diameter culets and Re gaskets were used to contain the sample. Neon was loaded as the PTM. Pressures were determined in situ from the using the ruby scale [18]/ equation of state of Cu standard (JCPDS 04-0836) at the same position where the PXRD data were taken on the sample. The 2-D diffraction images were integrated using the DIOPTAS software [19] and Rietveld and LeBail refinements were performed in GSAS-II [20].

### Synchrotron Mössbauer spectroscopy (SMS)

High-pressure $^{151}Eu$ SMS experiments were carried out at 3-ID-B Beamline of the APS, at ANL. SMS, also known as nuclear forward scattering (NFS), utilizes a pulsed synchrotron x-ray source to probe nuclear hyperfine interactions in the time domain rather than the energy domain used for conventional Mössbauer spectroscopy. The SMS experiments were performed in the 24-bunch timing mode with a 153 ns separation between successive electron bunches. A specially designed helium-flow cryostat cooled the sample to 50K while high pressures were generated using a membrane-driven miniature panoramic DAC [21]. 500 μm culet size diamonds were used as anvils and a laser drilled Re gasket formed the sample chamber. The SMS spectra were fitted using the CONUSS software package [22].

Both single crystal and sintered powder samples were measured by SMS. All the SMS measurements were performed at 50 K. This temperature was chosen for operational convenience: low enough to benefit from a factor of two increase in the recoil-free fraction ($f$-factor) compared with ambient temperatures [9] but high enough to remain above the expected pressure-driven increase in $T_N$. Helium was used as PTM to promote hydrostatic conditions at high pressure. After gas loading at room temperature, all subsequent pressure changes were made by the gas membrane at 50 K. Using an on-line system, laser excited ruby fluorescence spectra were collected for the determination of pressure using the ruby scale [18]. Pressure-induced changes in the europium valence, as well as structure, have been detected through either a change in the isomer shifts of the two pre-existing components ($Eu^{2+}$ and $Eu^{3+}$) or changes in their relative proportions. Absolute isomer shift values were obtained at several pressures by adding a reference sample (EuS) with a known isomer shift (-11.496 mm/s relative to $EuF_3$) in the x-ray beam [22,24-26].

### Partial fluorescence-yield X-ray absorption spectroscopy (PFY-XAS)

PFY-XAS experiments were carried out at the 16ID-D Beamline of the APS, at ANL, to provide direct information about the europium valence state and corroborate any possible changes suggested by the isomer shift measurements. The XAS experiment was carried out at Eu $L_3$ edge (6.97 keV, $2p_{3/2} \rightarrow 5d$ transition) at pressures up to ~36 GPa. A pair of 300 μm culet diamonds were used as anvils. The sintered powder sample (a small piece in ~ 50 μm size) was loaded together with a tiny ruby sphere into the aperture of the laser-drilled beryllium gasket and an insert formed by cubic boron nitride and epoxy. No PTM was added with the sample serving as its own transmission medium. As a result, the PFY-XAS measurements are expected to suffer the most from potentially non-hydrostatic effects. Pressures were measured in situ using ruby fluorescence [18]. To avoid heavy absorption by the diamond anvils at these low (~7 keV) x-ray energies, the XAS data were taken with the incident beam going through the beryllium gasket and the absorption signal being taken in fluorescence geometry (90° to the incident beam) using a Pilatus detector. The X-rays were focused to 5 μm (the Full Width Half Maximum (FWHM)). The beam spot location was carefully determined by scanning the sample position to minimize self-absorption.

**Acknowledgments**

Work at Ames National Laboratory is supported by the US DOE, Basic Sciences, Material Science and Engineering Division under contract no. DE-AC02-07CH11358. T.J.S. was partially supported by the Center for Advancement of Topological Semimetals (CATS), an Energy Frontier Research Center funded by the U.S. Department of Energy Office of Science, Office of Basic Energy Sciences, through Ames National Laboratory. Work at McGill University is supported by Fonds Québécois de la Recherche sur la Nature et les Technologies, and the Natural Sciences and Engineering Research Council (NSERC) Canada; Work at Argonne National Laboratory is supported by the U.S. Department of Energy, Office of Science, under contract No. DE-AC-02-06CH11357. Portions of this work were performed at GeoSoilEnviroCARS (The University of Chicago, Sector 13), Advanced Photon Source (APS), Argonne National Laboratory. GeoSoilEnviroCARS is supported by the National Science Foundation – Earth Sciences (EAR – 1634415). This research used resources of the Advanced Photon Source, a U.S. Department of Energy (DOE) Office of Science User Facility operated for the DOE Office of Science by Argonne National Laboratory under Contract No. DE-AC02-06CH11357. Use of the COMPRES-GSECARS gas loading system was supported by COMPRES under NSF Cooperative Agreement EAR -1606856 and by GSECARS through NSF grant EAR-1634415 and DOE grant DE-FG02-94ER14466. This research used resources of the Advanced Photon Source, a U.S. Department of Energy (DOE) Office of Science User Facility operated for the DOE Office of Science by Argonne National Laboratory under Contract No. DE-AC02-06CH11357. Portions of this work were performed at HPCAT (Sector 16), APS, ANL. HPCAT operations are supported by DOE-NNSA's Office of Experimental Sciences. GCJ and WB acknowledge the support from the National Science Foundation (NSF) CAREER Award No. DMR-2045760, and the help from Y. Xiao at APS, ANL, to set up the beamline. W. X. and H-Z W are supported by the U. S. Department of Energy (DOE), Office of Science, Basic Energy Sciences under award DE-SC0023648. PCC would like to thank Richard Olson for having helped enable the completion of this work.

**Figures and Tables**

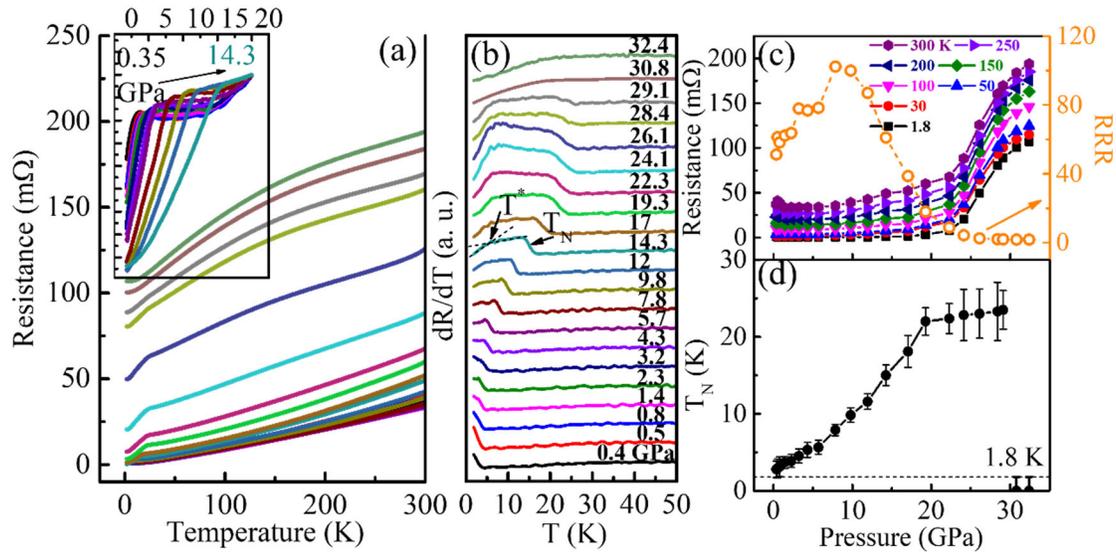

**Figure 1. Pressure dependence of electrical resistance and T_N.** Temperature dependencies of the (a) electrical resistance and (b) temperature derivatives of the resistance for single crystal EuPd$_3$S$_4$, measured at different pressures. (a) and (b) share the same legend. The inset in (a) shows the zoomed-in resistance curves normalized at 20 K from 0.4 GPa to 14.3 GPa so as to better show the evolution of T$_N$ below 14.3 GPa. In (b), curves at various pressures are offset from each other so as to better show the evolution of T$_N$. T$_N$ is determined as the midpoint of the dR/dT shoulder, and T* is defined as the intersection of two black dashed lines drawn from high temperature and low temperature. (c) Resistance at various temperatures and the residual resistivity ratio (RRR) as functions of pressure. (d) T$_N$ as a function of pressure. The dashed line denotes 1.8 K, which is the lowest temperature of these measurements.

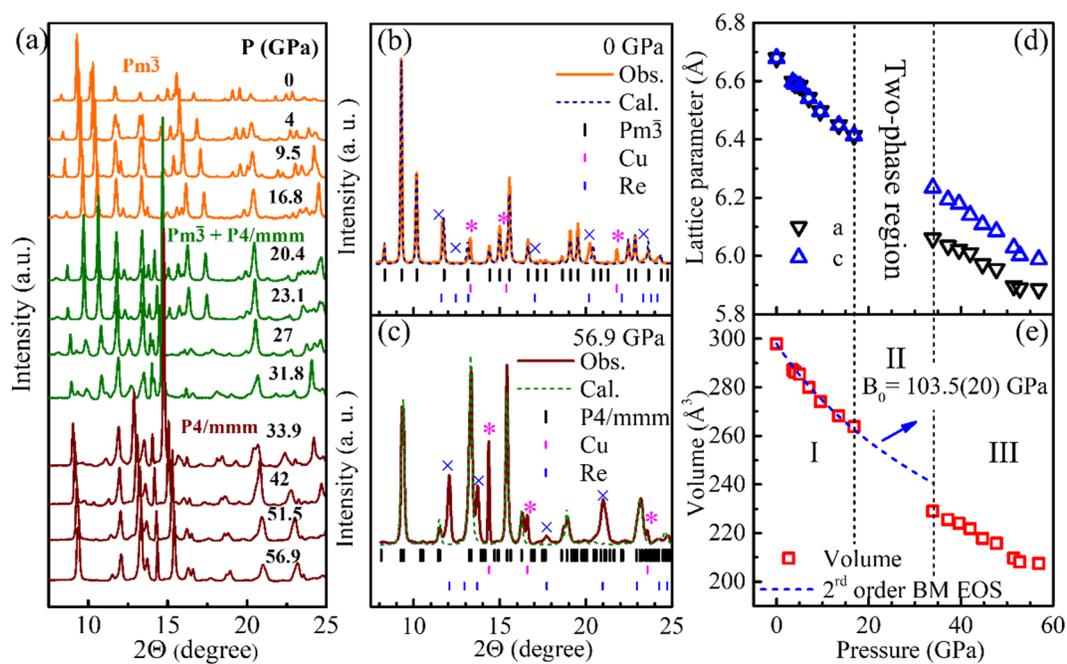

**Figure 2. Pressure induced structural phase transition.** (a) Selective PXRD patterns of EuPd$_3$S$_4$ at various pressures from 0 GPa to 56.9 GPa. The synchrotron x-ray wavelength λ is 0.4833 Å. The data was collected at room temperature. The full set is shown in Fig. S1. (b) Rietveld refinement of the PXRD pattern at ambient pressure. (c) Le Bail fit to the high-pressure P4/mmm structure at 56.9 GPa. The pink stars and blue crosses in (b) and (c) represent the clear extra peaks in the PXRD spectra of copper (pressure monometer) and rhenium (gasket), respectively. (d) Pressure dependences of the refined lattice constants a, and c. (e) Pressure dependence of the unit cell volume. The blue dashed curve shows the fitting by second-order Birch-Murnaghan equation of state (BM EOS). Several sets of independent PXRD measurements in zone I have been performed and show the same evolution of volume with the pressure. A more detailed plot is shown in Fig. S5.

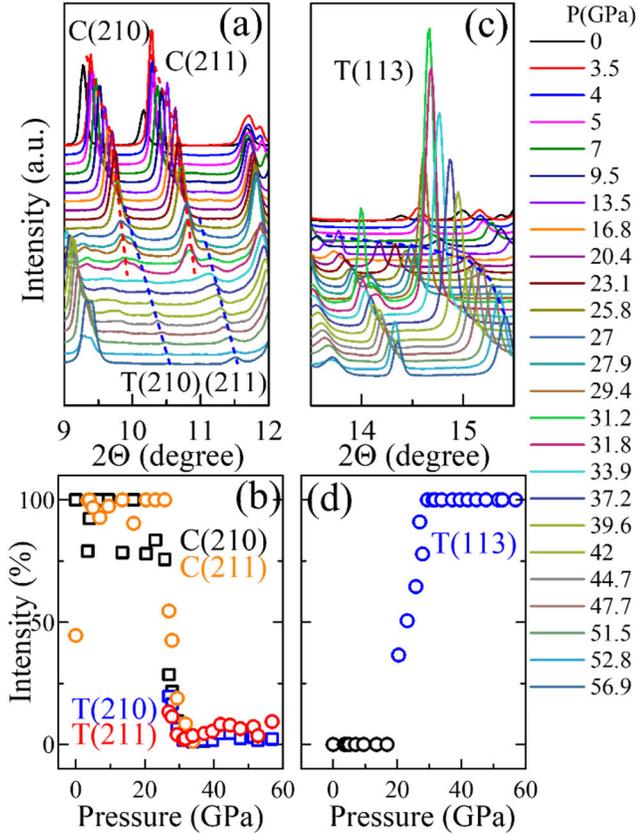

**Figure 3. The evolution of diffraction peaks with the pressure in selected 2θ ranges.** (a), (b) the cubic phase (210) and (211) ("cubic" is identified by "C") to tetragonal (210) and (211) ("tetragonal" is identified by "T"); (c), (d) the emergence of T(113). (a) and (c) share the same legend. The red and blue dashed curves in (a) and (c) are the smooth guides to the eye of the position change of diffraction peaks. The percentage of the intensities in (b) and (d) are estimated by: (intensity of peak)/ (maximum peak intensity in the spectrum).

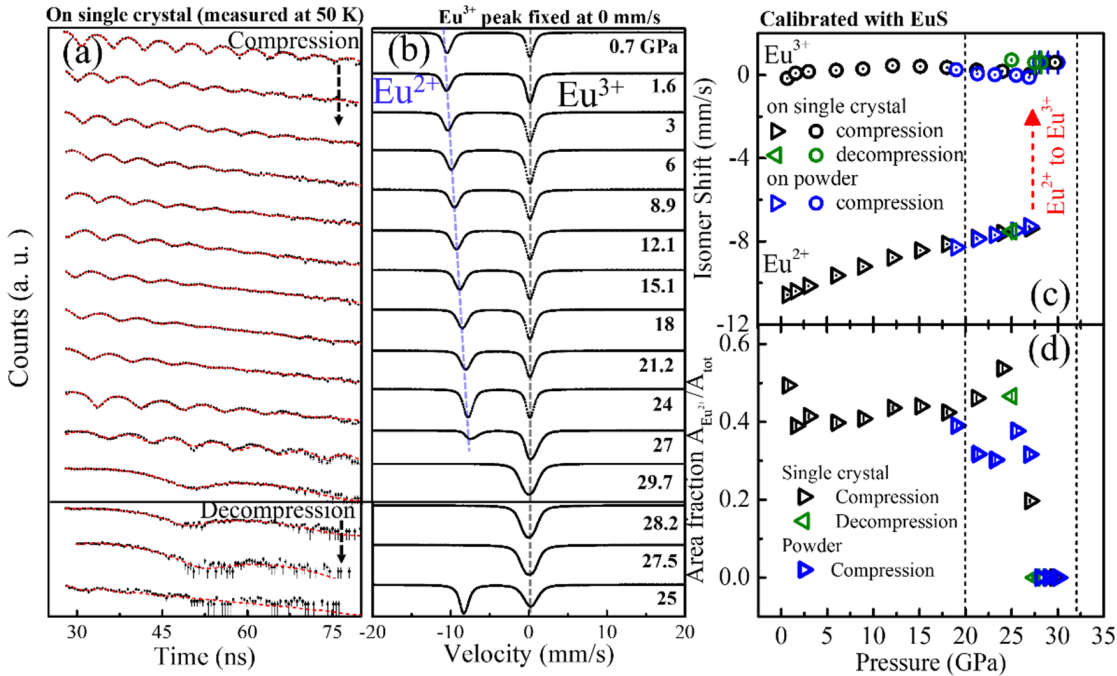

**Figure 4. Pressure-induced Eu valence transition from the isomer shift measurements.** (a) The time domain $^{151}$Eu Synchrotron Mössbauer spectroscopy spectra (SMS) of EuPd$_3$S$_4$ at 50 K under various pressures taken using a single crystal sample with helium as the pressure transmitting medium. The pressure was measured at 50 K. Pressure was first increased from 0.7 GPa to 29.7 GPa, then decreased to 25 GPa, in order to assess the reversibility of the pressure and the valence transition. The red dotted curves are fit to the SMS spectra using CONUSS [22]. (b) Simulated energy-domain Mössbauer spectra based on the fitting results from each corresponding SMS spectrum. For these simulated patterns, the isomer shift of the Eu$^{3+}$ peak was fixed to zero and only changes in the isomer shift and weight of the Eu$^{2+}$ peaks are apparent. In (b), the blue and black dashed lines serve as visual aids to illustrate the evolution of Eu$^{2+}$ isomer shift while Eu$^{3+}$ remains fixed. (a) and (b) share the same legend. The calibrated (using the EuS standard) pressure dependence of the absolute isomer shifts for the Eu$^{2+}$ and Eu3$^+$ peaks are shown for both samples in (c), while (d) shows the pressure dependence of the Eu$^{2+}$ area fraction. The absolute value of isomer shift is calibrated by moving a reference sample (EuS, at room temperature and ambient pressure, with known isomer shift of Eu$^{2+}$ at -11.496 mm/s) on the beam line. The detailed calibration process is explained in Fig. S8, S9.

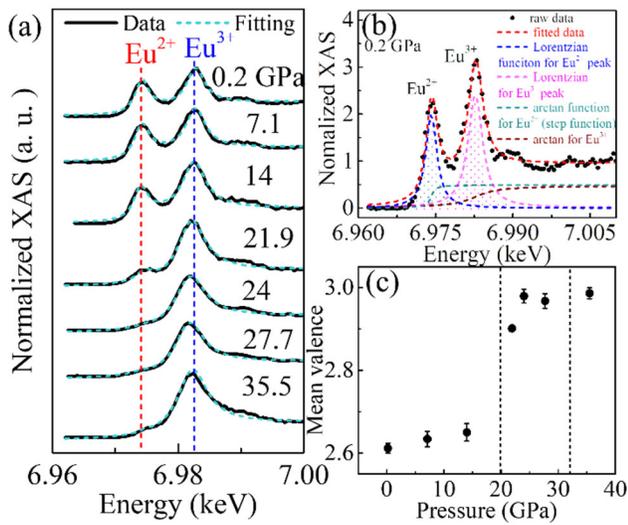

**Figure 5. Pressure induced valence transition from absorption spectra**. (a) PFY-XAS data at Eu $L_3$ absorption edge in $EuPd_3S_4$ at room temperature at pressures from 0.2 GPa to 35.5 GPa showing the valence transition from roughly 50:50 $Eu^{2+}$:$Eu^{3+}$ to essentially $Eu^{3+}$. The red and blue vertical dashed lines are guides to the eye showing the absorption peak energies for $Eu^{2+}$ and $Eu^{3+}$, respectively. (b) Fits to the 0.2 GPa PFY-XAS data (described in the text). The red dashed line represents the total summation of 2 arctangent step functions and 2 Lorentzian peaks. (c) Mean Eu valence as a function of the pressure. The mean valence is estimated by from the ratio of the areas of fitted Lorentzian peaks.

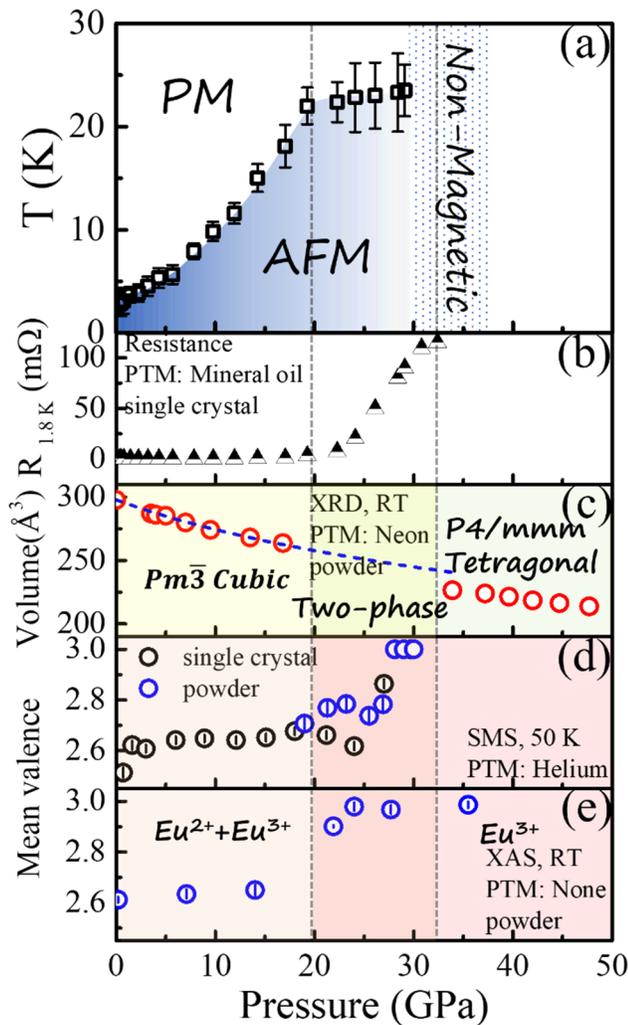

**Figure 6. Pressure dependence of electronic, magnetic, and structural properties of EuPd$_3$S$_4$.** The pressure dependence of (a) AFM transition temperature T$_N$, (b) resistance at 1.8 K, (c) unit cell volume estimated from the high-pressure X-ray diffraction measurement, (d) mean valence estimated from the Synchrotron Mössbauer spectroscopy (SMS) spectra, and (e) mean valence estimated from the Partial fluorescence-yield X-ray absorption spectroscopy (PFY-XAS) spectra. PTM: pressure transmitting medium. RT: room temperature. The blue dashed curve in (c) is the second-order BM EOS fitted curve that shows the clear difference in the unit cell volume between zone I and zone III. The vertical dashed lines crossing all panels mark the lower and upper boundaries of the two-phase region determined from x-ray diffraction.

## Supporting Information Text

### Synchrontron X-ray diffraction

As clearly shown in Fig. S2 (a), the relative intensities of (210) and (211) peaks of the cubic phase ("cubic" is identified as "C") are stable in Zone I, and strongly suppressed in Zone II, accompanied by the emergence of extra peaks at a little higher angle than (210) and (211), which could be (210) and (211) peaks in tetragonal phases ("tetragonal" is identified as "T"). In Zone III, C(210) and C(211) peaks are not detectable anymore, whereas T(210) and T(211) survive with very low relative intensities. The diffraction peaks' positions and relative intensities of (200) and [(210)&(211)] as a function of pressure are plotted in Fig. S2 (b), (c) and (d), (e), respectively. The relative peak intensities are in percents and are obtained by first removing the background and diffraction peaks of copper and rhenium from the spectra and then normalizing the spectra to the diffraction peak with the highest intensity. The peak positions of (200), (210) and (211) in both cubic or tetragonal phases show almost linear behavior as a function of pressure in Zones I and III. Above ~26 GPa, the peak position of (200) shows a small jump to higher angle, and a second jump happens just at ~34 GPa. On the other hand, C(210) and C(211) coexist with T(210) and T(211) peaks in Zone II, and the peaks' positions are almost independent of pressure. The different evolution of peak position in Zone II indicates the possible existence of strain effect between two distinct phases or/and the different compressibility of different phases. As for the peaks' intensities, C(210) and C(211) decrease rapidly from almost 100 % to 0 in zone II, whereas T(210) and T(211) peaks decrease from ~ 25 % to less than 10 % in Zone II and become almost constant. Although there appears to be an increase in intensity in zone III, however, the systematic evolution of (200) peak with pressure is difficult to observed because C(200) peak actually splits into 2 peaks T(200) and T(002) under high pressure and these two peaks are very close to each other, as shown in the inset of Fig. S2 (c). More typical case of peak splitting will be discussed later.

In Fig. S3 (a), we show that a new peak with the peak position around 14.2° (T(113) according to our indexing) emerges at ~20 GPa and systematically shifts to the higher angle with pressure increase. The peak position evolves linearly with the pressure above ~32 GPa, and deviates a little bit from a linear relation in the pressure region between ~20 GPa and ~32 GPa, shown in Fig. S3 (b). The intensity of T(113) increases with the pressure rapidly from ~20 GPa to ~32 GPa and it becomes the strongest diffraction peak at higher pressure, as shown in Fig. S3 (c).

Fig. S4 (a) shows that the C(400) peak splits into two peaks i.e. T(400) and T(004) above 27 GPa. A mixed phase is also observed in Zone II, where we can see several small peaks merging together at around the angle of C(400). The accurate information of peak positions and intensities of C(400), T(400) and T(004) could not be obtained in Zone II, whereas in Zone III, clear and well separated T(400) and T(004) peaks are observed. The peak splitting also happens to lower angle diffraction peaks that have small splitting angle such as in C(200). The peak splitting under high pressure suggests that the structure has changed to lower symmetry which is consistent with the suggestion of a cubic to tetragonal structural transition.

### Synchrotron Mössbauer spectroscopy

**Way to calibrate the absolute value of isomer shifts with EuS as reference.** As shown in Fig. S9, at several pressure points, the reference sample EuS, at room temperature and ambient pressure was placed together with EuPd$_3$S$_4$, which is at 50 K. The collected time-domain spectra were then fitted with CONUSS [22] by fixing the isomer shift of Eu$^{2+}$ site of EuS as -11.496 mm/s, and varying isomer shifts and their weights of all Eu sites of EuPd$_3$S$_4$ (2 sites below 27 GPa, and 1 site at and above 27 GPa). Since Eu$^{2+}$ peak (as demonstrated in Fig. 4(b)) is more sensitive to pressure. The black circles are the fitted points of isomer shift of Eu$^{2+}$ calibrated by EuS. The black circles can be well fitted with an exponential function shown with the red dashed curve. The detailed fitted parameters are shown in the inset. According to the fitting, we can estimate that the isomer shift of Eu$^{2+}$ is about ~10.75 mm/s at ambient pressure, which is supposed to be the isomer shift of Eu$^{2+}$ and is very close to the reported value (~10.9 mm/s) [9]. The time-domain spectra, collected without reference sample, were fitted by fixing the isomer shift of Eu$^{3+}$ site of EuPd$_3$S$_4$ as 0 mm/s, and varying isomer shift and the weight of Eu$^{2+}$ sites of EuPd$_3$S$_4$ (below 27 GPa). The blue circles are the relative isomer shifts of Eu$^{2+}$ fitted with this method. Based on the exponential fitting, we suppose that the absolute isomer shifts of Eu$^{2+}$ at different pressures will drop on the red curve. The isomer shifts of Eu$^{3+}$ at different pressures, then, should be equal to the vertical differences between the red curve and blue circles, as Δδ, shown in the figure. The detailed plot of the pressure dependence of absolute isomer-shift values is shown in Fig. 4(c).

**Way to estimate the mean valence.** The mean valence of the sample at various pressures is usually estimated simply by the area of energy domain absorption peaks, based on the assumption that the area of the absorption peaks is proportional of the number of Eu ions in particular valence. The formula we use is:

$$v = \frac{2 \times A_{Eu^2} + 3 \times A_{Eu^3}}{A_{Eu^2} + A_{Eu^3}} \quad (1)$$

where v is the mean valence, A is the area.

It is worth noting that the valence fluctuation between the $4f^7$ and $4f^6$ configurations typically occurs on the order of $10^{-11}$ s [24]. However, Mössbauer spectroscopy operates on a characteristic time scale of approximately $10^{-8}$ s, which is three orders of magnitude longer than the valence fluctuation time. As a result, Mössbauer spectroscopy may only yield an average isomer shift from the intermediate valence state instead of distinct isomer shifts for $Eu^{2+}$ and $Eu^{3+}$ when the isomer shift of $Eu^{2+}$ begins to shift towards higher energies, indicating an oxidation state approaching 3+. Given that there is already an approximate 50:50 $Eu^{2+}$:$Eu^{3+}$ at ambient pressure, which is the $EuPd_3S_4$ case, we may observe both separated Eu3+ isomer shift and an average Eu isomer shift under pressure (represented by the $Eu^{2+}$ peak in Fig. 4 and Fig. S6). And using the empirical formula 4, the average isomer shift δ for very short fluctuation times could be expressed by:

$$\delta = p_{Eu^{2+}}\delta_{Eu^{2+}} + p_{Eu^{3+}}\delta_{Eu^{3+}} \quad (2)$$

Where $p_{Eu^{2+}}$ and $p_{Eu^{3+}}$ are the relative populations of the 2+ and 3+ states. $\delta_{Eu^{2+}}$ and $\delta_{Eu^{3+}}$ are the absolute isomer shift of 2+ and 3+ state. Based on the consistent evolution of isomer shift as a function of the pressure in Fig. 4, it is convenient to fix $\delta_{Eu^{3+}}$ at zero and use the linear fit to estimate the $\delta_{Eu^{2+}}$ and $\delta_{Eu^{3+}}$. Then the average of valence $v_-$ is given by:

$$v_- = 3 - \frac{\delta}{\delta_{Eu^{2+}}} \quad (3)$$

Then we can improve the formula (1) as:

$$v = \frac{v_- \times A_{Eu^2} + 3 \times A_{Eu^3}}{A_{Eu^2} + A_{Eu^3}} \quad (4)$$

Fig. 6 (d) shows the mean valence as a function of the pressure after fixing. The estimation is made under the assumption that the pressure effect is small.

**SI References**

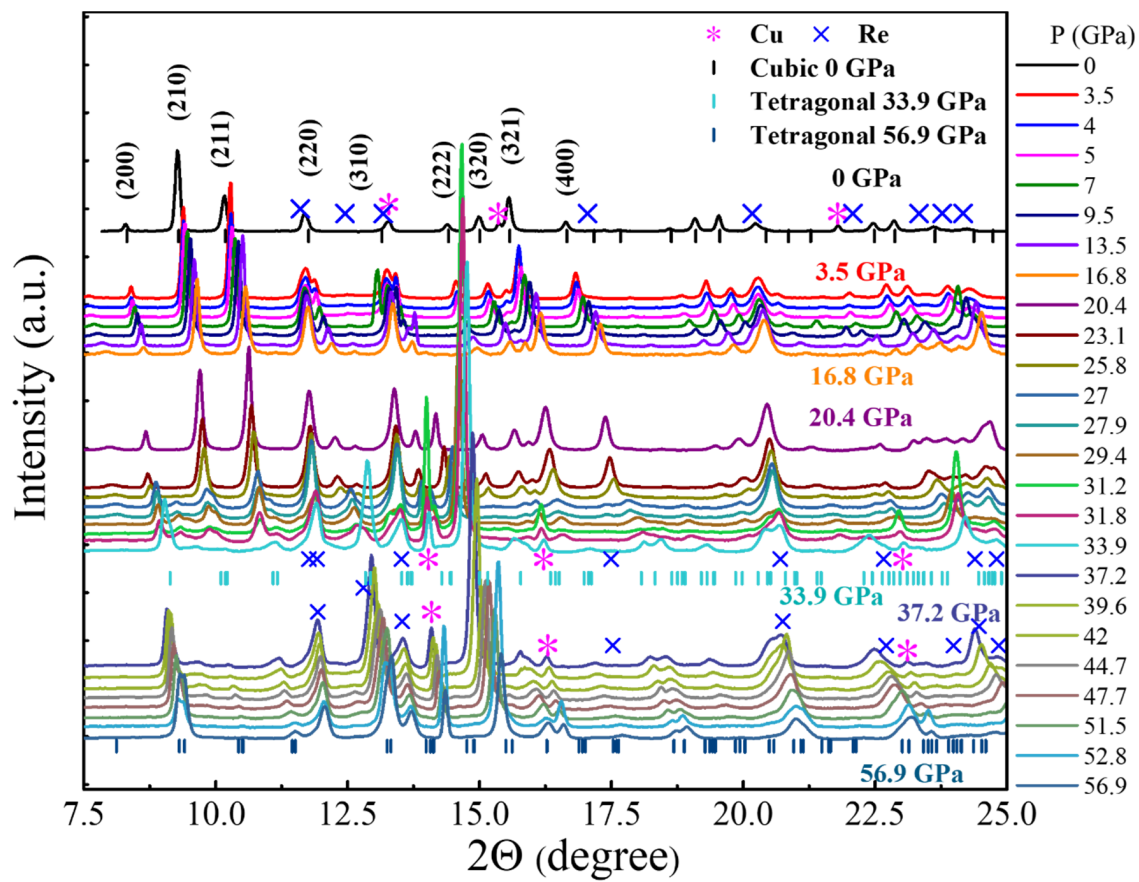

**Fig. S1. Pressure induced the structural phase transition.** The PXRD patterns of EuPd$_3$S$_4$ at various pressures from 0 GPa to 56.9 GPa. The synchrotron x-ray wavelength λ is 0.4833 Å. The data are collected at room temperature. The pink stars and blue crosses represent peaks in the PXRD spectra of copper and rhenium, respectively.

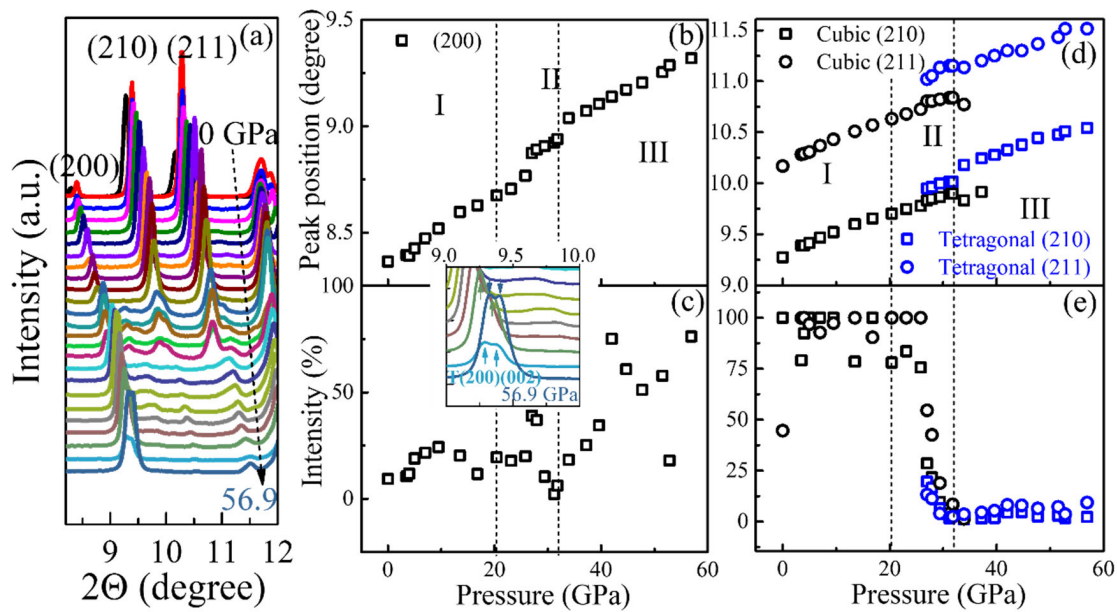

**Fig. S2. Demonstration of typical cases for structural phase transition under high pressure in selected 2θ ranges.** (a) Showing the evolution of (200), (210) and (211) peaks with the pressure. (a) uses the same legend as Fig. S1. Pressure increases from top (0 GPa) to bottom (56.9 GPa). (b) and (d) Diffraction peaks positions of (200), (210) and (211) as a function of pressure. (c) and (e) Relative intensity of (200), (210) and (211) peaks as a function of pressure. The inset in (c) is a zoomed view of the pattern showing the peak splitting from cubic (200) to tetragonal (200) and (002). The pressure region between the two dashed lines corresponds to the two-phase region where both cubic and tetragonal phases are present.

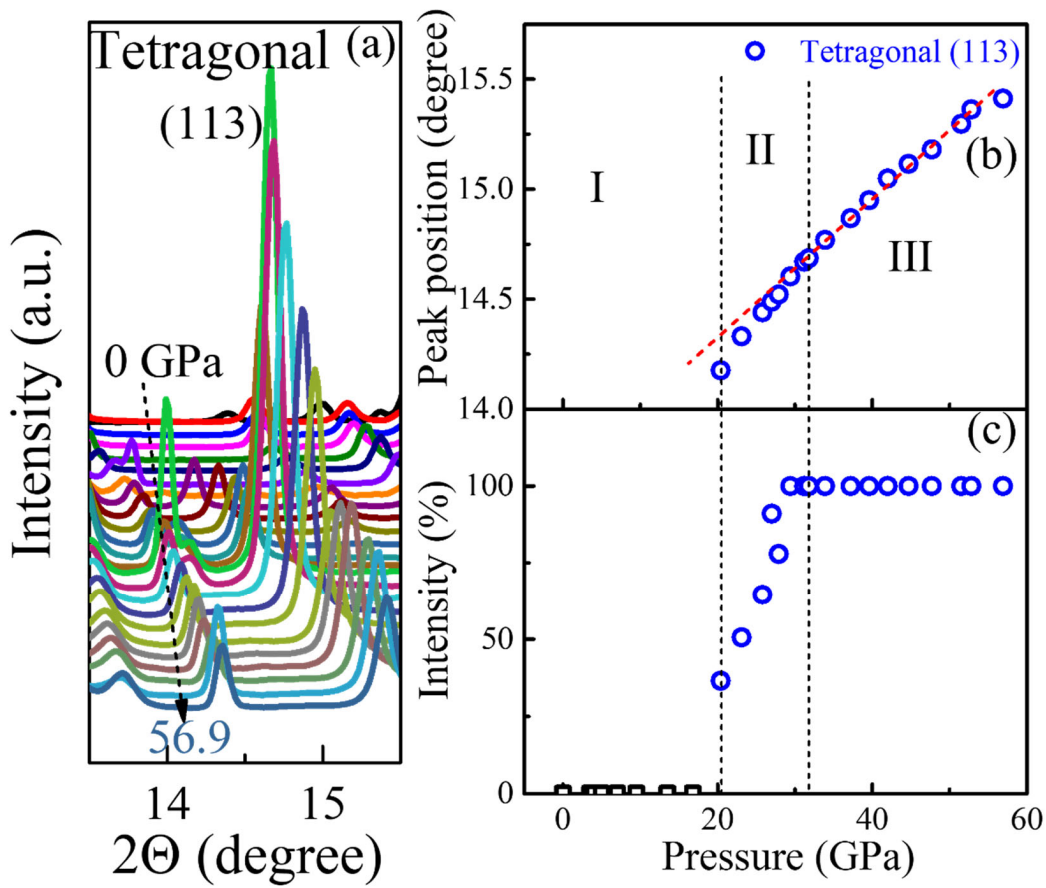

**Fig. S3. Demonstration of typical cases for structural phase transition under high pressure in selected 2θ ranges: new peak emergence.** (a) The new peak (possible tetragonal (113)) appears under high pressure. (a) uses same legend as Fig. S1. Pressure increases from top (0 GPa) to bottom (56.9 GPa). (b) Diffraction peak position of possible (113) in tetragonal phase as a function of pressure. (c) Relative intensity of (113) peak as a function of pressure. The pressure region in between two dashed lines might be in a two-phase region of the cubic phase and tetragonal phase. The red dashed straight line is a guide to the eye.

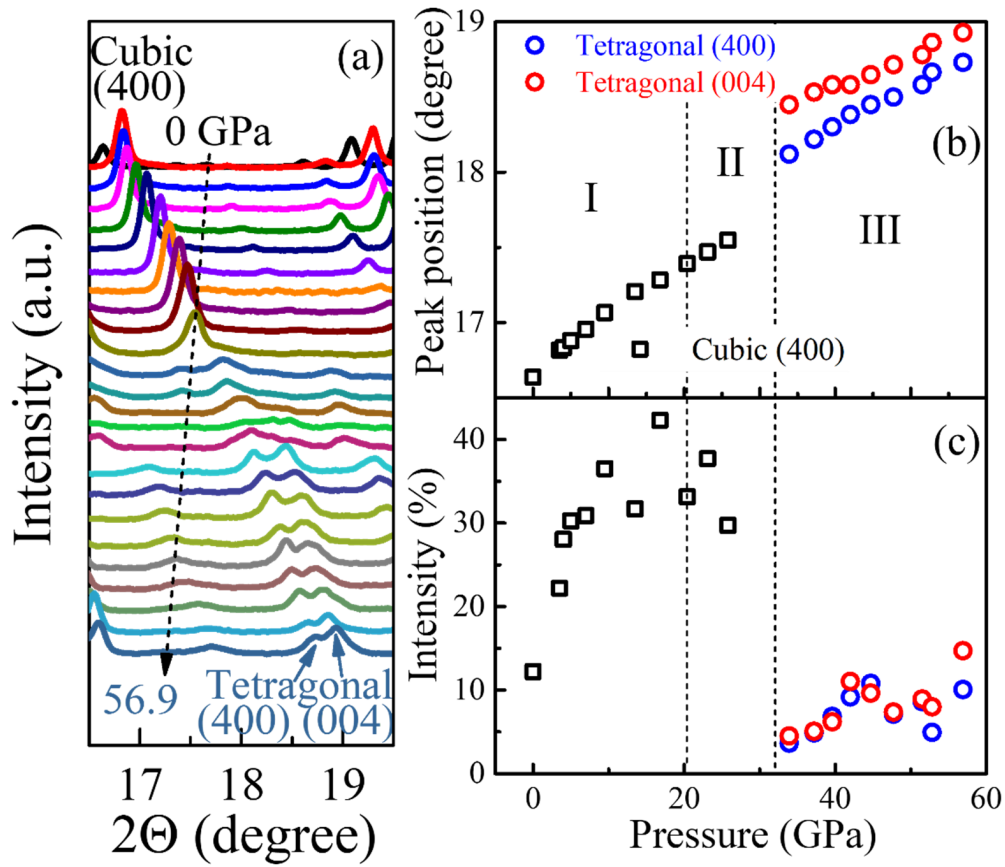

**Fig. S4. Demonstration of typical cases for structural phase transition under high pressure in selected 2θ ranges: peak splitting.** (a) The cubic (400) Bragg peak splits into tetragonal (400) and (004) peaks under high pressure. (a) uses same legend as Fig. S1. Pressure increases from top (0 GPa) to bottom (56.9 GPa). (b) Diffraction peak positions of the cubic (400) peak and the tetragonal (400) and (004) peaks as a function of pressure. (c) Relative intensities of the cubic (400) and tetragonal (004) peaks as function of pressure. The pressure region in between two dashed lines contains a mixture of cubic and tetragonal phases.

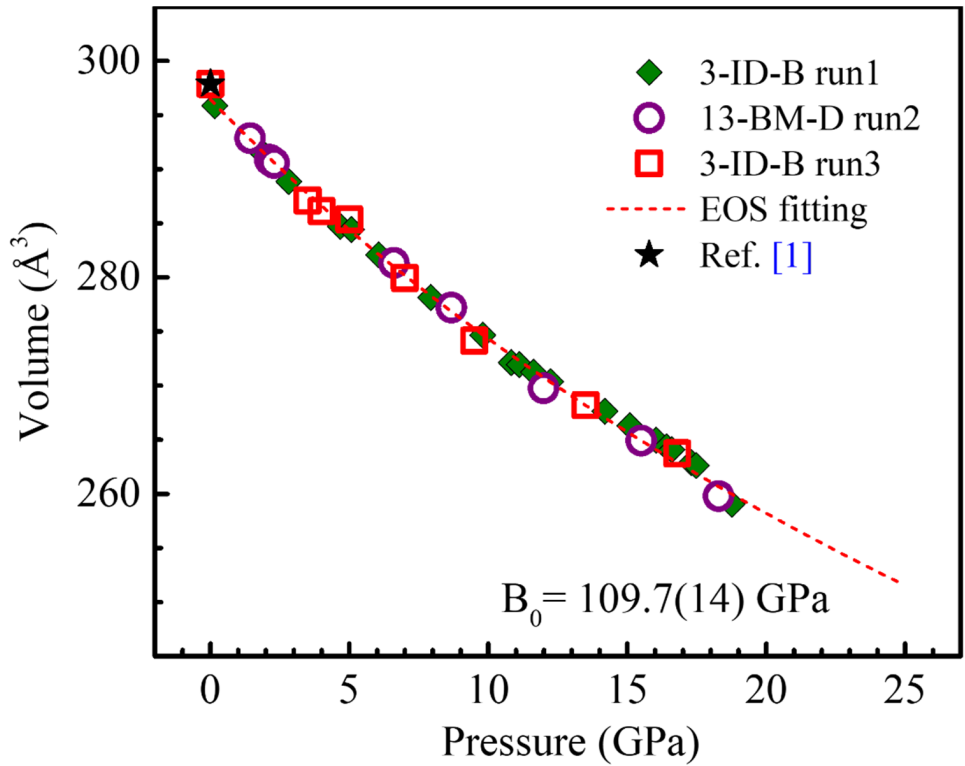

**Fig. S5. The consistency in volume evolution with pressure in zone I.** The unit cell volume as a function of pressure, based on several independent runs of PXRD measurements (run1 to run3). All PXRD data are put together and could be fitted well with a second-order BM EOS (shown by red dashed curve), with zero-pressure volume $V_0$= 296.54 Å$^3$, which is close to the previous report value (~297.9 Å$^3$) [9], and zero-pressure bulk modulus $B_0$=109.7(14) GPa. This figure clearly shows the agreement between different runs.

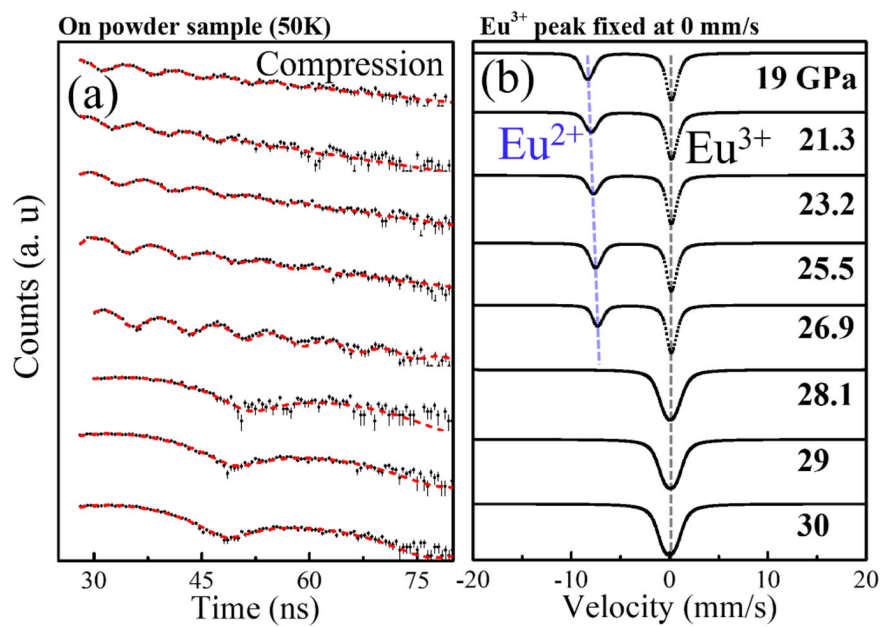

**Fig. S6. Pressure induced Eu valence transition from the isomer shift.** (a) The Synchrotron Mössbauer spectroscopy spectra (SMS) in time domain of $^{151}$Eu from EuPd$_3$S$_4$ at 50 K under different pressures on sintered powder sample. The red curves show the fit of the SMS spectra. (b) Simulated energy-domain Mössbauer spectra based on the fitting results from each corresponding SMS spectrum. (a) and (b) share the same legend.

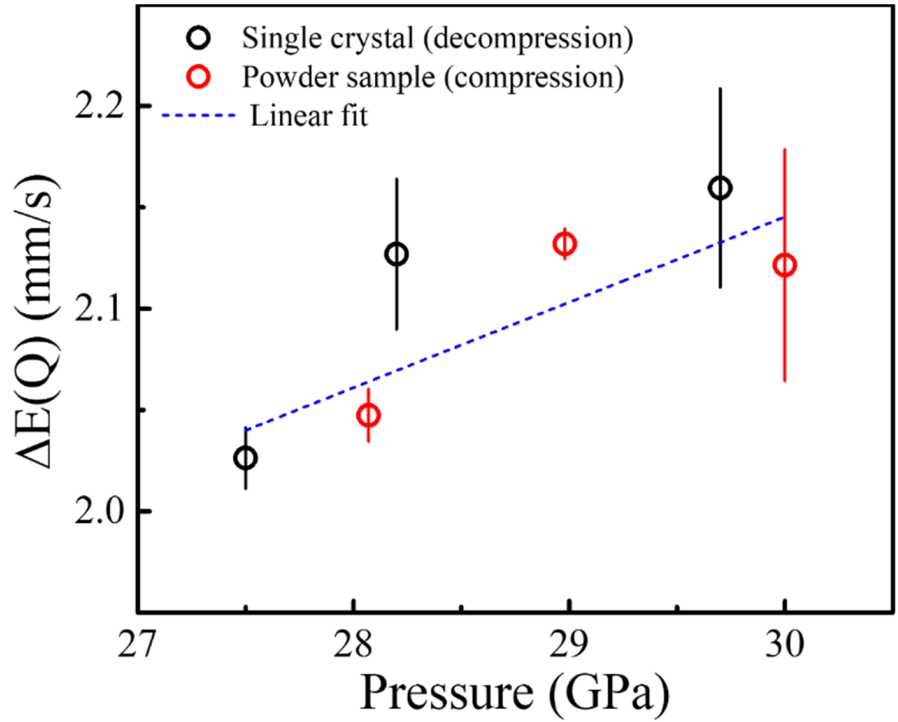

**Fig. S7. Pressure dependence of quadrupole interaction, ΔE(Q).** When the valence transition occurs, a distinct change was observed in the time-domain Mössbauer spectra, indicating the disappearance of $Eu^{2+}$. The spectra can be accurately fitted by introducing a quadrupole splitting term at the $Eu^{3+}$ site, attributed to the presence of an electric field gradient in the lower symmetry, tetragonal unit cell. As such then, this observation aligns effectively with the structural transition from cubic to tetragonal. Meanwhile, it is noted that the quadrupole interaction tends to have small amount of increase with the pressure at the range between ~28 and ~30 GPa.

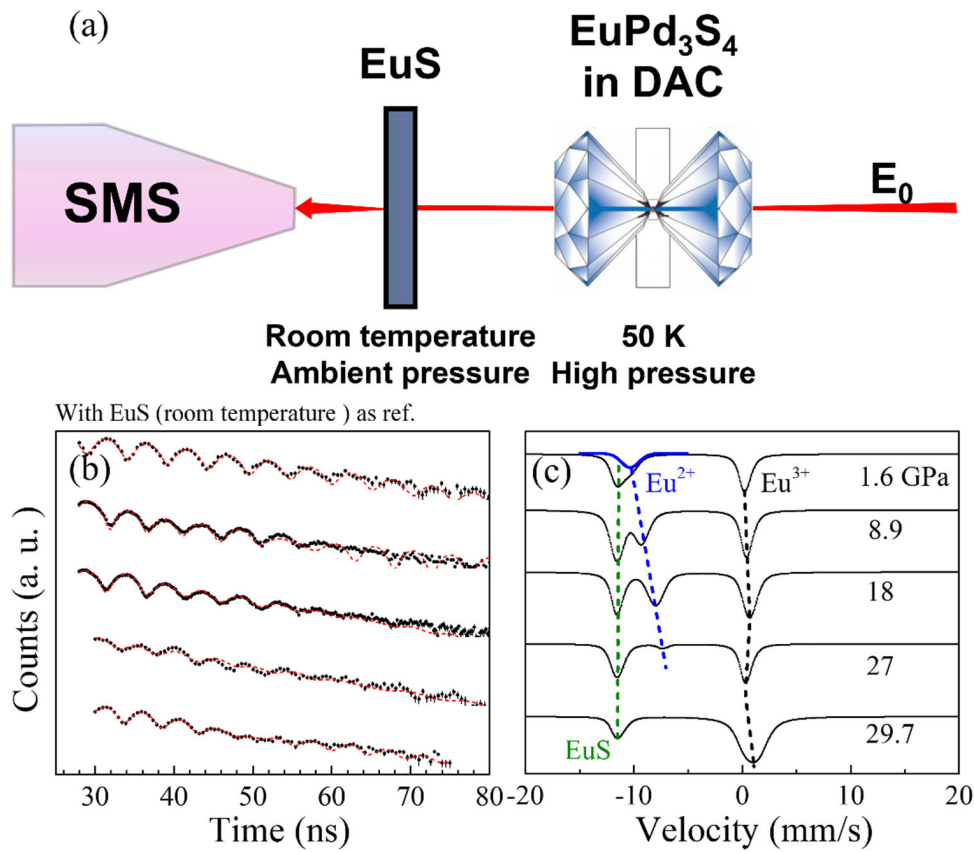

**Fig. S8. Calibration of the absolute value of isomer shifts with EuS as reference.** (a) Schematic drawing of the SMS measurement setup with reference sample EuS. (b) The SMS spectra in the time domain of $^{151}$Eu from EuPd$_3$S$_4$ at 50 K at various pressures, obtained from experimental measurements conducted on a single crystal sample. To determine the absolute value of the isomer shift and its variation with applied pressure, a reference sample of EuS is employed at room temperature. The red dashed curves represent the fittings of the SMS spectra. (c) Energy-domain spectra simulations are performed based on the fitted results obtained from each corresponding time-domain SMS spectrum. The green dashed line indicates the isomer shift of Eu$^{2+}$ of the reference sample EuS, which remains fixed at -11.496 mm/s. The blue and black dashed lines serve as visual guides illustrating the evolution of the Eu$^{2+}$ and Eu$^{3+}$ peaks of EuPd$_3$S$_4$, respectively, as pressure increases. This depiction agrees with the results depicted in Fig. 4(b). The blue Gaussian peak observed just above 1.6 GPa corresponds to the fitted peak of the Eu$^{2+}$ peak in EuPd$_3$S$_4$. Both (a) and (b) share the same legend.

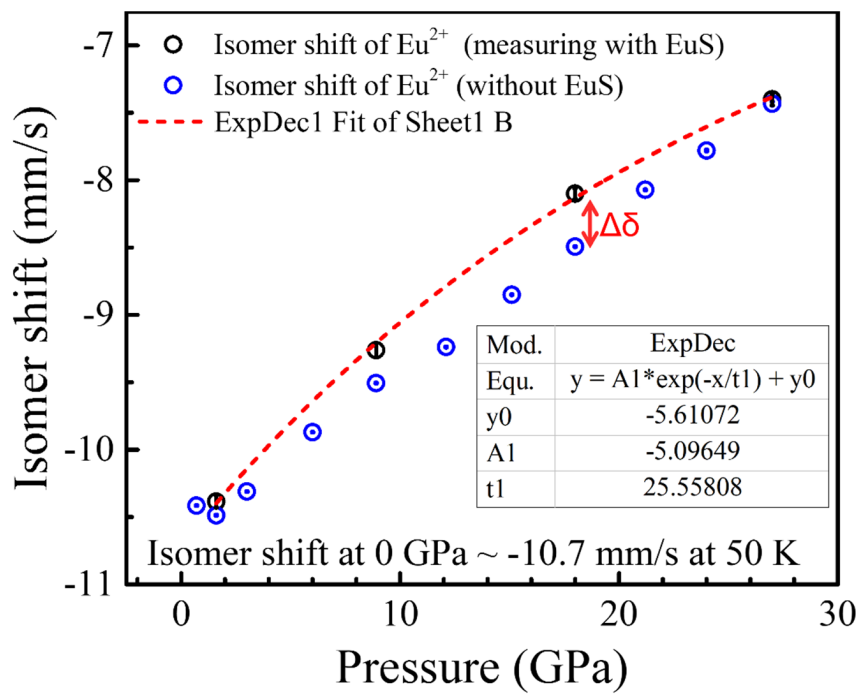

**Fig. S9.** Calibration of the absolute value of isomer shifts with EuS as reference.

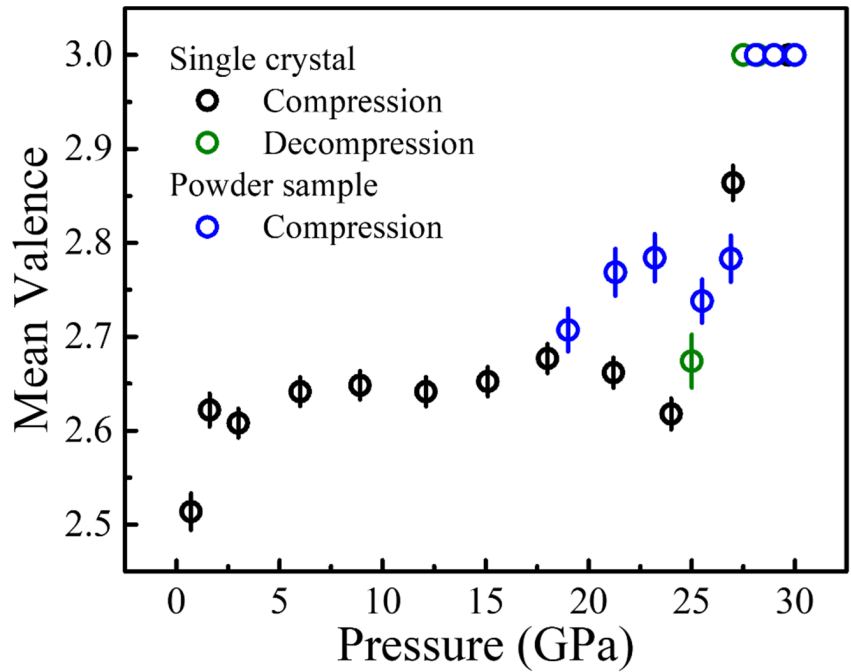

**Fig. S10.** Mean valence estimated by the areas of the absorption peaks assuming that above the critical pressure of valence transition, the absorption peak with negative isomer shift is assumed to be an average peak with partial $Eu^{2+}$ and $Eu^{3+}$ due to the valence fluctuation.